\documentclass[11pt]{amsart}
\usepackage{geometry}                
\usepackage{graphicx}
\usepackage{amssymb}
\usepackage{epstopdf}
\usepackage{longtable}
\usepackage{threeparttable}
\usepackage{booktabs}
\usepackage{psfrag}
\usepackage{overpic}
\usepackage{color}
\usepackage{caption2}
\usepackage{subfigure}

\title{A heat production model for stable continental lithosphere by the inversion of the surface heat flows}
\author{Rong Qiang Wei}
\address{College of Earth Sciences, University of Chinese Academy of Sciences, Beijing, PRC, 100049}
\email{wrq1973@ucas.ac.cn}
\date{}

\begin{document}
\maketitle
\begin{abstract}

Obtaining the heat production (HP) in the lithosphere has always been a challenge for the geotherms and evolution of continents. By transforming the nonlinear stable heat conduction equation into a linear Poisson's potential one, we propose a method to infer the HP in the stable continental lithosphere. This method estimates the HP through the inversion of the corresponding heat flow (HF) observations. Either the distribution of HP within the lithosphere or geometry of HP interfaces (even the both) can be inverted.  Herein we focus on the inversion of the geometry of the HP interface.  An analogical Parker-Oldenburg formula, which is often used in the inversion of potential field, is deduced for this purpose. This analogical formula is based on a relationship between the Fourier transform of the corresponding HF observations and the sum of the Fourier transform for the powers of the interface geometry. When the mean depth of the HP interface and the HP contrast between the two media are given, one to three-dimensional geometry of the HP interface can be iteratively calculated. As a case study, we construct a HP model for the lithosphere of the Ordos geological block and its adjacent area in China, in which three-dimensional geometry of the upper crustal HP interface is estimated. It is found that the geometry of this HP interface distribute uneven in a magnitude of dozens of km.  With this HP model, the geotherms for the Ordos geological block and its adjacent area is calculated further, which fit the constraints from the studies on the xenolith and tectonics well. 

\end{abstract}

{\hspace{2.2em}\small Keywords:}

{\hspace{2.2em}\tiny Heat production,  Geotherms, inversion, Stable continental lithosphere, the Ordos geological block}

\section{Introduction}\label{intro}

Heat production (HP) is one of the key parameters for the calculation of the geotherm of the continental lithosphere. Previous studies showed that HP accounts for more than 30\% of heat loss through the continents (e.g., Pollack and Chapman, 1977; Artemieva and Mooney, 2001; Hasterok and Chapman, 2011).  Therefore it is necessary and important to obtain the distribution of the HP within the continental lithosphere. However, by common geophysical and geochemical exploration methods, it is difficult to obtain measurements of HP at depths greater than a few hundred meters. The scientific deep drilling holes, e.g. Kola (Russia, Popov et al., 1999), KTB (Germany. Clauser et al., 1997), Cajon Pass (USA. Williams et al., 1988), and CCSD (China. He et al., 2008), provide direct measurements of HP at deeper depths, but there is only a few and sparse deep drillings in the world. These facts result in that HP is one of the least constrained physical parameters in the continental lithosphere.

Except scientific deep drilling holes, there are two other "direct" approaches to estimate the distribution of the HP in the lithosphere. One is to study the HP distribution in the titled section of the deep crust that exposed on the surface. For example, Vredefort basement in Kaapvaal craton (South Africa), exposes almost 15 km thick vertical section of the Archean crystalline crust due to overturning of Vredefort granitic basement because of palaeo-geotectonic processes. It offers an opportunity of studying the deep crustal HP (Hart et al., 1981). Similar section includes Northeastern Baltic shield (Europe. Kremenetsky et al., 1989), Hidaka Metamorphic belt (Hokkaido, Japan. Furukawa and Uyeda, 1989), Nissho Pluton, Hidaka Terrain (Hokkaido, Japan, Furukawa and Uyeda, 1989), and Dharwar Craton (Southern India, Ray et al., 2003). Another is to study simultaneously the HP and geobarometry data on rocks, in which the space information is inferred from the geobarometry data (eg., Brady et al., 2006).  Likewise, they are limited to specific areas. 

To obtain the distribution of the HP in the lithosphere, lots of indirect methods have been adopted. Some authors attempted to estimate the HP from the seismic velocity structure using the empirical relationship between the heat production value and seismic velocity (Rybach and Buntebarth, 1982, 1984; \v Cerm\'ak and Rybach, 1989). This approach provides the possibility of constructing a three-dimensional HP distribution in the lithosphere through the lithospheric three-dimensional seismic velocity structure. But because of the large scatter of heat production values for the rocks with the same seismic velocities, and the great non-uniqueness of the HP of any given rock, the empirical relationship has large errors (Kremenetsky et al., 1989). 

Because of the difficulty in determining the three-dimensional HP distribution in the lithosphere, a typical HP model in which the HP varies with depths, is often used as both a reference and/or a starting model to calculate the geotherm of the continental lithosphere. Some of these HP models are estimated from the linear relationship observed between HF and HP at the surface (Birch et al., 1968; Roy et al., 1968; Lachenbruch, 1968; Turcotte and Schubert, 2014). A step, a linear decreasing, an exponential decreasing, and a modified exponential decreasing model of HP were suggested from this relationship (e.g., Lachenbruch, 1970; Zang et al., 2002, 2005). The exponential model is considered to be the preferable because it is the only one that preserves such a linear HF-HP relation under differential erosion (Lachenbruch, 1970), and this exponential depth dependence is also consistent with magmatic and hydrothermal differentiation precesses (Turcotte and Schubert, 2014). But none of the published data from boreholes penetrating deeply into the crystalline continental crust show clear evidence for a systematic variation of heat production with depth, neither linear nor exponential (Clauser et al, 1997).

With the accumulation of the observation data and measurements, many other HP models for the crust or the lithosphere were suggested. Based on the variation in crustal HP with crustal age,  a HP model for the crust was presented (Jaupart and Mareschal, 2003).  Another HP model estimated upper crustal HP from surface HP measurements constrained by a heat budget (e.g., Chapman,1986; Artemieva and Mooney, 2001), or from xenolith P-T conditions (Rudnick et al., 1998; Rudnick and Nyblade, 1999). Moreover, HP distribution was presumed based on the petrological structural model for the lithosphere together with HP measurements (e.g., Kremenetsky et al., 1989; Chi and Yan, 1998). Vil$\acute{a}$ et al. (2010) studied the variability of some common lithological groups from a compilation of a total of 2188 measurements, and presented a layered HP model for a standard crustal column according to Wedepohl (1995). Hasterok and Chapman (2011) proposed a layered HP model for continental lithosphere, based on the petrology of the lithosphere, HP measurements of surface and xenolith samples, and tectono-thermal constraints. Huang et al. (2013) constructed a reference Earth layered model for the HP constrained by geoneutrino experimental results. This is a new way to obtain the distribution of the HP in the lithosphere. More and detailed HP models can be found in a recent review by Jaupart et al. (2016) which summarized information extracted from large data sets on HP.

\begin{figure}[htb]
\setlength{\belowcaptionskip}{0pt}
\centering
\begin{overpic}[scale=0.55]{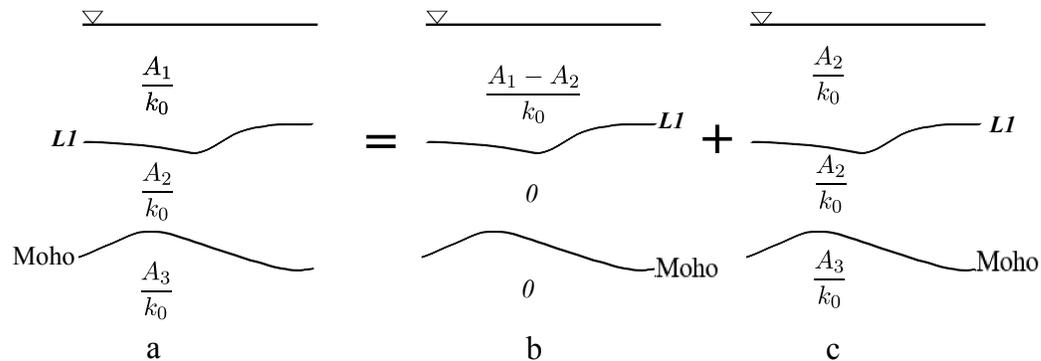}
\end{overpic}
\renewcommand{\figurename}{Fig.}
\caption{\footnotesize (a) The sketch map of a three-layered heat production (HP) model for the stable continental lithosphere. The geometry of $L1$ is to be inferred by inversion of the surface HF. For this purpose, model (a) is then decomposed into a superposition of model (b) and (c). $A_1$, $A_2$, $A_3$ are the HP of the upper crust, lower crust, and lithospheric mantle, respectively. $k_0$ is the coefficient of thermal conductivity measured at a reference temperature $T_0$. Moho is the interface between the lower crust and lithospheric mantle.}
\label{fig1}
\end{figure}

The HP models above are deterministic function. However, on the other hand, some researchers represented the HP as a stochastic quantity.  For example, Nielsen (1987) assumed that HP can be stated as the sum of an average term and a small, zero mean perturbation, when he studied the heat conduction in a heterogeneous medium. Srivastava and Singh (1998) took HP as a known mean value and a Gaussian correlation structure when they studied the temperature variations in sedimentary basins. 

Geoscientists get much benefit from these HP models to estimate the geotherms of the lithosphere. A reasonable HP model for the continental lithosphere should fit both the observation of the surface HF and the measurements of the HP of the rocks at least. Here we propose such a probable layered HP model by the inversion of the measurements of surface HFs.  The related theory and methodology will be given in section \ref{sec2}. Because this method is only available for stable continental lithosphere, the Ordos geological block and its adjacent area is taken as a case study. Its HP model for the lithosphere is constructed in section \ref{sec3}. Based on this HP model, in the section \ref{sec4} we calculated the geotherms of the lithosphere for this geological block and its adjacent area. Some conclusions are drawn in the section \ref{sec5}.

\begin{figure}[htb]
\setlength{\belowcaptionskip}{0pt}
\centering
\begin{overpic}[scale=0.5]{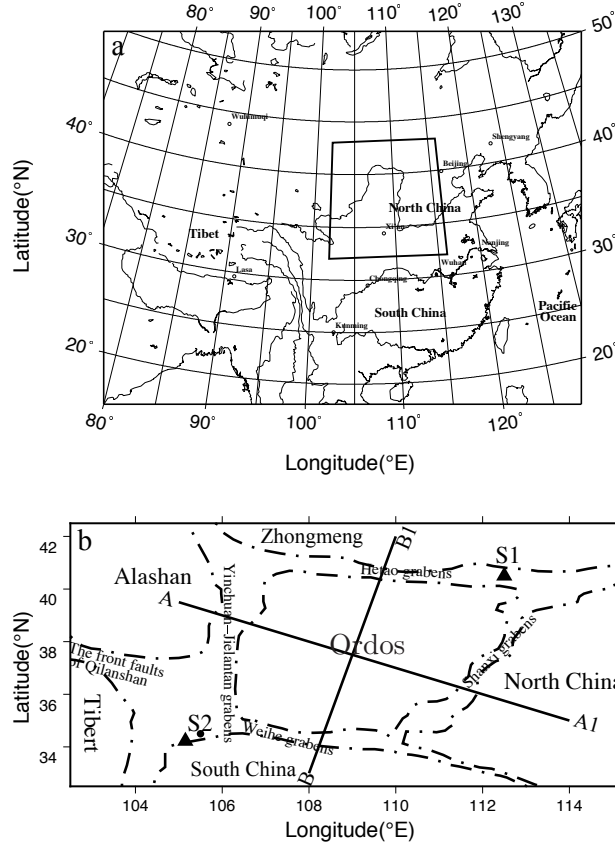}
\put(37.5,22.5){Ordos}
\end{overpic}
\renewcommand{\figurename}{Fig.}
\caption{\footnotesize (a) Location of the Ordos geological block and its adjacent area in China (the area shown within a box). (b)Map of the study area. The names of the geological blocks are shown in bold fonts. The dash-dotted line are boundary of the geological blocks. The Ordos block is separated by four grabens which consist mainly of faults in different geology age (Zhang et al. 2003). Two solid lines are the location of the cross-sections studied in this paper. Two black triangles are the locations of the xenolith outcrop. S1 ($112.5^\circ\rm{E}, 40.5^\circ\rm{N}$) and S2 ($105.5^\circ\rm{E}, 34.5^\circ\rm{N}$) are the two sites where the one-dimensional geotherm will be calculated.}
\label{fig2}
\end{figure}

 \section{Theory and Methodology}\label{sec2}
 \subsection{Theory}
 
 We start from the three-dimensional equations of the steady heat conduction,
 
\begin{equation}\label{eq1}
\nabla\cdot[(k(T) \nabla T]+A=0
\end{equation}

where $T=T(x,y,z)$ is the temperature, $k(T)$ the coefficient of thermal conductivity which depends only on the $T$, $A=A(x,y,z)$ the HP.

By Kirchoff transform as follows,
 
 \begin{equation}\label{eq2}
 U(T)=\int_{_{T_0}}^{T}\frac{k(T')}{k_0}{\rm{d}}T'
 \end{equation}

 where $T_0$ is a reference temperature, and $k_0$ the coefficient of thermal conductivity measured at $T_0$.
 
We can transform equation (\ref{eq1}) into (\ref{eq3})
 
 \begin{equation}\label{eq3}
\nabla^2 U+\frac{A}{k_0}=0
\end{equation}

\begin{figure}[htb]
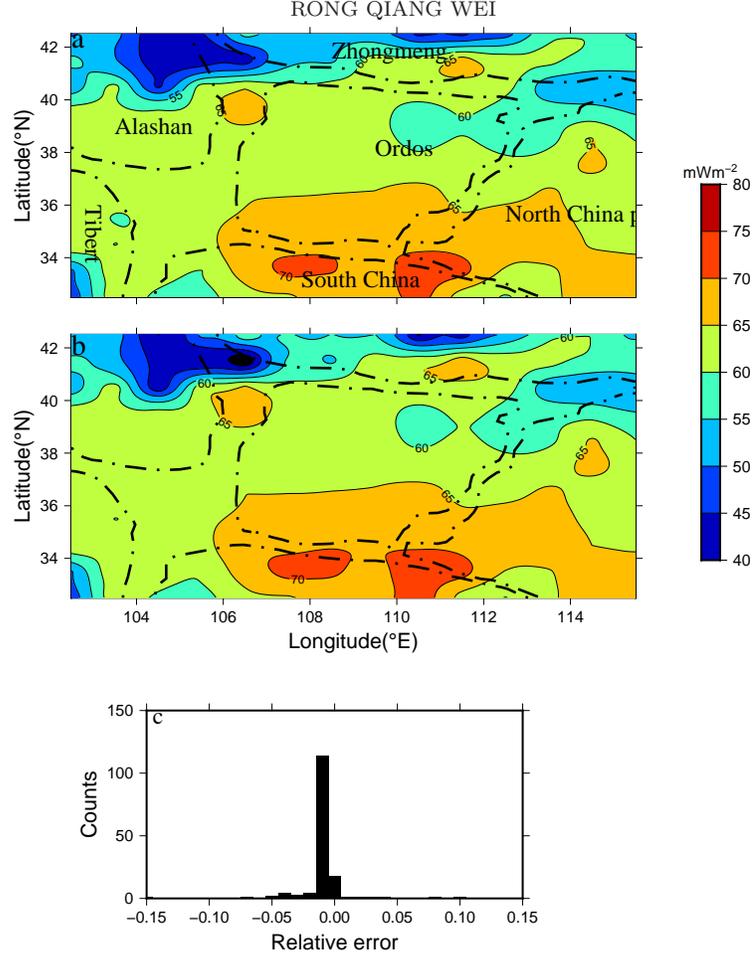

\setlength{\belowcaptionskip}{0pt}
\centering
\begin{overpic}[scale=0.5,bb=15 57 581 751]{Hf_ob_cal.eps}
\end{overpic}
\renewcommand{\figurename}{Fig.}
\caption{\footnotesize (a) Surface HF for the Ordos geological block and its adjacent areas (Data is from Goutorbe et al.(2011)) (b) Surface HF calculated from the HP model obtained in this paper (Figure \ref{fig1}a). (c) The histogram of relative error.}
\label{fig3}
\end{figure} 

Equation (\ref{eq3}) is Poisson's equation. A special solution $U(\textbf{r})$ of this equation is,

\begin{equation}\label{eq4}
\begin{array}{ll}
U(\textbf{r}) =U(x,y,z) &=\frac{1}{4\pi k_0}\int_V\frac{A(x',y',z') \mbox{d}V}{\vert \textbf{r}-\textbf{r}_0\vert} \\
   &=\frac{1}{4\pi k_0}\int_V\frac{A(x',y',z') \mbox{d}V}{[(x-x')^2+(y-y')^2+(z-z')^2]^{1/2}}
\end{array}
\end{equation}

where $\textbf{r}$ is the radius vector of the field point $(x,y,z)$,  $\textbf{r}_0$ is that of the source point $(x',y',z')$, $V$ the volume of the entire HP distribution.

Further, we can get (\ref{eq5}) from (\ref{eq4}),

\begin{equation}\label{eq5}
\frac{\partial U}{\partial z}=-\frac{1}{4\pi k_0}\int_V\frac{A(x',y',z')(z-z') \mbox{d}V}{[(x-x')^2+(y-y')^2+(z-z')^2]^{3/2}}
\end{equation}

Moreover, according to Kirchoff transform, i.e., equation (\ref{eq2}), we have,

\begin{equation}\label{eq6}
\begin{array}{ll}
 \frac{\partial U}{\partial z} &=\frac{\partial U}{\partial T}\frac{\partial T}{\partial z}\\
  &=\frac{k(T)}{k_0}\frac{\partial T}{\partial z}\\
  &=-\frac{q}{k_0}
  \end{array}
\end{equation}

where $q$ is the HF.

At the surface $z=0$, $q=-q_0$ where $q_0$ is the surface HF, and we can get,

\begin{equation}\label{eq7}
\frac{\partial U}{\partial z}\vert_{_{z=0}}=\frac{q_0}{k_0}
\end{equation}

It can be seen from equation (\ref{eq5}) and  (\ref{eq7}) that $A$ can be estimated by the inversion of the measurements of $q_0$.

\begin{figure}[htb]
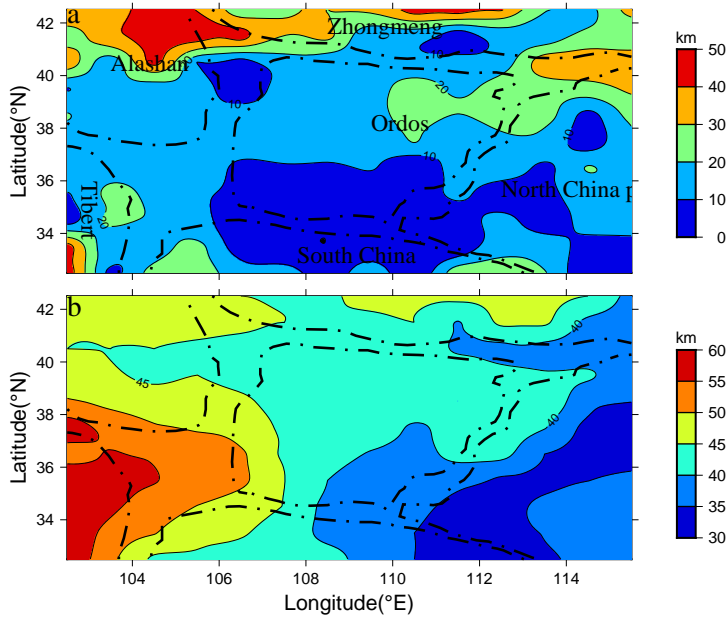

\setlength{\belowcaptionskip}{0pt}
\centering
\begin{overpic}[scale=0.5,bb=7 278 582 750]{Depth_hprod2.eps}
\end{overpic}
\renewcommand{\figurename}{Fig.}
\caption{\footnotesize (a) Depths for $L1$ for the Ordos geological block and adjacent areas, which are inferred from the inversion of the corresponding variation of the surface $q_0/k_0$. (b) The depths of the Moho discontinuity from CRUST1.0 model (Data is from Laske et al. (2013)), which are used directly the HP interface between the lower crust and the lithospheric mantle.}
\label{fig4}
\end{figure}

\subsection{Methodology}

For simplicity, herein we assume a HP model like that shown in Figure \ref{fig1}a. This HP model consists of three layers: the upper crust (HP is $A_1$), the lower crust ($A_2$),  and the lithospheric mantle ($A_3$); $L1$ and Moho are the upper-lower crustal HP  interface, and the lower crust-lithospheric mantle HP interface, respectively. The bottom of this HP model is at infinity downward.

Like the problem in the usual inversion of potential field, the variation of the surface $q_0/k_0$ can be attributed to $A_1/k_0$, $A_2/k_0$ and $A_3/k_0$, or the geometry of $L1$ and Moho, or both above. At present we have no a good method to distinguish between them. Here we assume that the second case holds, namely, the variation of the surface $q_0/k_0$ is only due to the geometry of $L1$ and Moho, and the HP within the interfaces is constant, i.e., $A_1$, $A_2$, $A_2$, $k_0$ are constant. Further, we will only infer the geometry for $L1$ from the corresponding variation of the surface $q_0/k_0$. While the Moho will be replaced directly by the Moho discontinuity from the seismology. The reason is that the upper crustal HP has a significant effect on the geotherm of the lithosphere, and the deviations from average lower crustal and lithospheric mantle HP can be easily masked by upper crustal HP uncertainties (Hasterok and Chapman, 2011). 

In the inversion of potential (gravity or magnetic) field, there are many methods to obtain the geometry of physical interface from the anomalies observed. Here we use a method similar to the Paker-Oldenburg approach (Parker, 1973; Oldenburg, 1974). Parker (1973) presented an expression that the Fourier transform of the anomalies observed is the sum of the Fourier transforms of the powers of the physical interface causing the anomalies. Oldenburg (1974) rearranged the Parker’s expression to determine the geometry of the physical interface by iteration. 

\begin{figure}[htb]
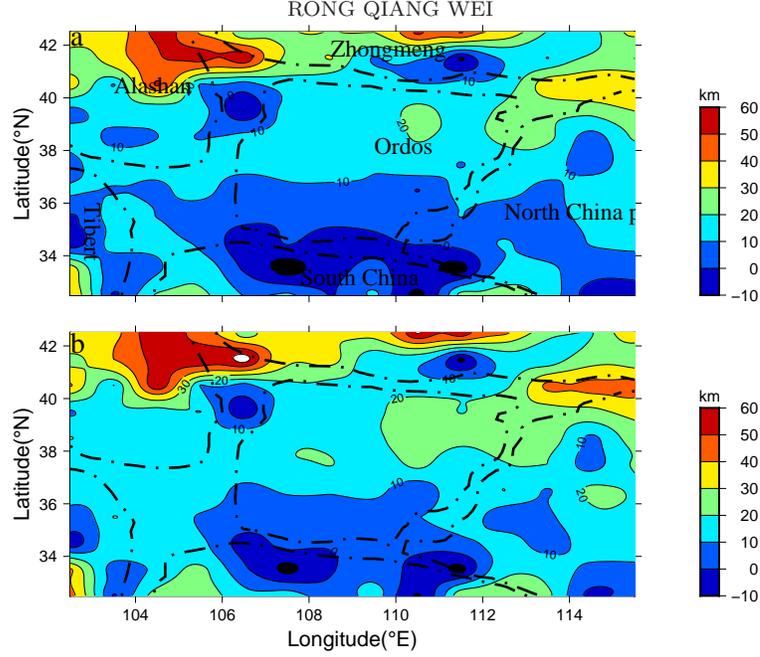

\setlength{\belowcaptionskip}{0pt}
\centering
\begin{overpic}[scale=0.5,bb=7 278 582 750]{Depth_hprod_z0_diff_perc10.eps}
\end{overpic}
\renewcommand{\figurename}{Fig.}
\caption{\footnotesize The histogram of relative error. (a)$z_0$ has a change of -10\% relative to $z_0$ in Figure \ref{fig4}; (b) $z_0$ has a change of 10\%}
\label{fig5}
\end{figure}

Following the Parker's expression (Parker, 1973), we consider the $\frac{\partial U}{\partial z}$(=$-\frac{q}{k_0}$) from a layer of HP of constant $A$, whose lower boundary is the plane $z = 0$, and whose upper boundary is defined by the equation $z = h(\textbf{r})$. We have,

\begin{equation}\label{eq8}
F[\frac{\partial U}{\partial z}] =-\frac{A}{2k_0}\exp(-\vert\vec{k}\vert z_0) \sum_{n=1}^{\infty}\frac{\vert\vec{k}\vert ^{n-1}}{n!}F[h^n(\textbf{r})]
\end{equation}

where $F[\cdot]$ denotes the Fourier transform, $\vec{k}$ is the wave number, $h(\textbf{r})$ the depth to the interface (positive downwards) and $z_0$ often the mean depth of the $h(\textbf{r})$.

\begin{figure}[htb]
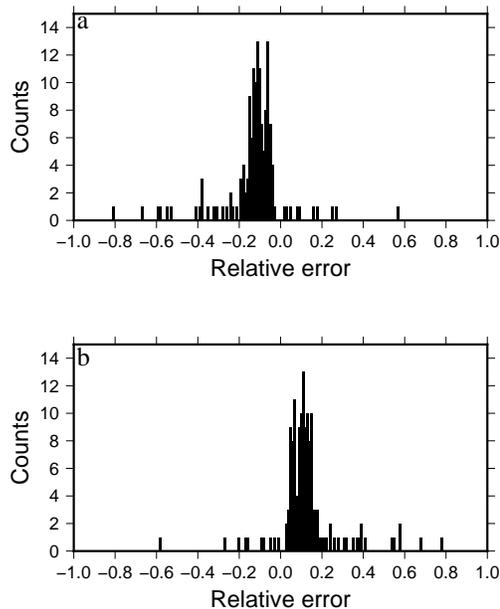

\setlength{\belowcaptionskip}{0pt}
\centering
\begin{overpic}[scale=0.55,bb=7 298 352 718]{HPdep_z0_diff_perc10.eps}
\end{overpic}
\renewcommand{\figurename}{Fig.}
\caption{\footnotesize Depths for $L1$ for the Ordos geological block and adjacent areas inferred from different $z_0$. (a)$z_0$ has a change of -10\% relative to $z_0$ in Figure \ref{fig4}; (b) $z_0$ has a change of 10\%.}
\label{fig5+1}
\end{figure}

To obtain the causative HP geometry or topography $h(\textbf{r})$, equation (\ref{eq8}) is rearranged according to the scheme of Oldenburg (1974) as follows,

\begin{equation}\label{eq9}
F[h(\textbf{r})]=-\frac{2k_0}{A}F[\frac{\partial U}{\partial z}]\exp(\vert\vec{k}\vert z_0)-\sum_{n=2}^{\infty}\frac{\vert\vec{k}\vert ^{n-1}}{n!}F[h^n(\textbf{r})]
\end{equation}

Equation (\ref{eq9}) allows us to determine the topography of the HP interface $h(\textbf{r})$ by an iterative inversion procedure, if the corresponding variation of the surface $q_0/k_0$ is known. 

Because Equation (\ref{eq9}) is only for computing the $q_0/k_0$ effect caused by a subsurface HP anomaly or topography, the model in Figure \ref{fig1}a should be decomposed into a superposition of model in Figure \ref{fig1}(b) and (c) to infer the geometry or topography of the $L1$. This decomposition is valid and reasonable because equation (\ref{eq3}) is linear. 

Thus, for the HP model in Figure \ref{fig1}, the outline to infer the geometry or topography of $L1$ is:

\begin{itemize}
\item Estimate the surface $\frac{q_0}{k_0}\vert_{_{\rm{Moho}}}$ caused by Moho using forward model (Equation (\ref{eq8}), and Figure \ref{fig1}c). 
\item Estimate the surface $\frac{q_0}{k_0}\vert_{_{L1}}$ generated from $L1$ by subtracting the $\frac{q_0}{k_0}\vert_{_{\rm{Moho}}}$ from the surface $\frac{q_0}{k_0}$ observed (Figure \ref{fig1}a).
\item Estimate the geometry of $L1$ by inversion of the $\frac{q_0}{k_0}\vert_{_{L1}}$ (Equation (\ref{eq9}), and Figure \ref{fig1}b) 
\end{itemize}

There are many open-source codes can be used for the forward and inverse modeling here based on Equation (\ref{eq9}) after a little modification. For example, the matlab code 3DINVER.M by G$\acute{\rm{o}}$mez-Ortiz and Agarwal (2005), or the forward and inverse Fortran programs by Shin et al. (2005). All of these codes are based on Parker-Oldenburg method and can be performed in three dimensional space. 

It should be pointed out that the methodology for a single layer here can be easily applied to multi-layer cases as well, but the HP model is more complicated than that in Figure \ref{fig1}.  On the other hand, there are also many ways to estimate the distribution of the $A/k_0$ (sources) by the inversion of the variation of the surface $q_0/k_0$, like inferring the distribution of the density or the magnetic intensity by inversion in the gravity and magnetic problem (e.g., Stocco et al., 2009; Camacho et al., 2011).

\begin{figure}[htb]
\setlength{\belowcaptionskip}{0pt}
\centering
\begin{overpic}[scale=0.45,bb=68 158 779 547]{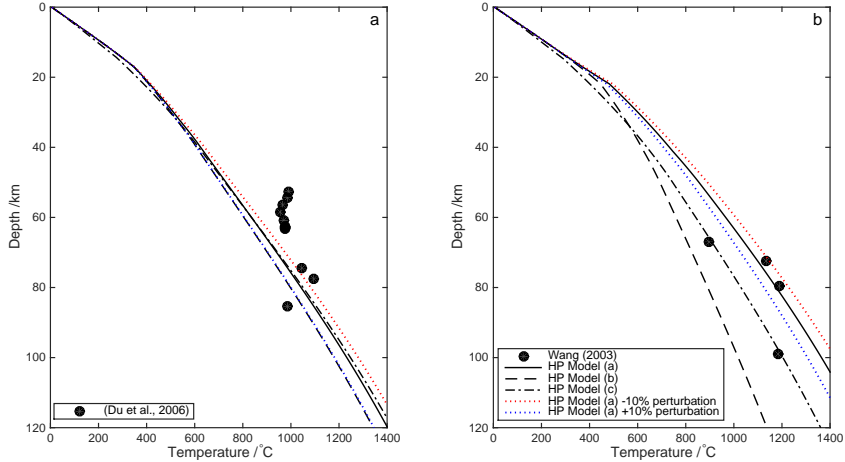}
\end{overpic}
\renewcommand{\figurename}{Fig.}
\caption{\footnotesize 1-D geotherm for two sites (a) S1 ($112.5^\circ\rm{E}, 40.5^\circ\rm{N}$); (b) S2 ($105.5^\circ\rm{E}, 34.5^\circ\rm{N}$). Solid circles in the figure are the temperature and pressure values for equilibrium mineral assemblages in cenozoic xenoliths discovered near S1 and S2, respectively.}
\label{fig6}
\end{figure}

\section{Geometry of $L1$ for the Ordos geological block and its adjacent areas}\label{sec3}

Because our start point is equation (\ref{eq1}), the methodology to infer the geometry of $L1$ described in section (\ref{sec2}) is only available to the stable areas. Here we apply this method to the Ordos geological block and its adjacent areas in China as an example ($102^\circ-116^\circ\rm{E}$,$32^\circ-43^\circ\rm{N}$), as shown in Figure \ref{fig2}.  The Ordos geological block is located between the Tibet plateau and the North China plain. There are no active tectonic events such as an uplift of the upper mantle, active Quaternary faults or large earthquakes within the Ordos geological block (Deng et al. 1999). Therefore the Ordos geological block is stable in tectonics and the methodology in section \ref{sec2} can be used.

\subsection{Data}

The Data required in the inversion are the following,

1. The surface HF

The surface HF is critical to the inversion of the geometry of $L1$. Although this inversion can be used in any scale, here we only infer the geometry of $L1$ on a grids of $1^\circ\times 1^\circ$, because the HF measurements are sparse and cover only a limited surface of the study areas. We use the surface HF distribution shown in Figure \ref{fig3}a for the Ordos geological block and its adjacent areas, which is a part of the global data set from Goutorbe et al. (2011). In this dataset the surface HF for grids without measurements are estimated based on multiple geological and geophysical proxies. Here for these grids, surface HF measured inferred from the similarity method are used,  for its better accuracy in cross-validation tests. To decrease the boundary effects, in the inversion we use a broader range ($100^\circ-120^\circ\rm{E}$,$30^\circ-50^\circ\rm{N}$) in which the Ordos geological block and its adjacent areas are included. 

2. The HP model and HP 

We use the HP model in Figure \ref{fig1}a, i.e., a three-layered one for the Ordos geological block and its adjacent area, in which the HP  is $A_1=1.50\rm{\mu Wm}^{-3}$ in the upper crust, $A_2=0.45\rm{\mu Wm}^{-3}$ in the lower crust, and $A_3=0.02\rm{\mu Wm}^{-3}$ in the mantle (Hasterok and Chapman, 2011; Furlong and Chapman, 2013).  In this model, the geometry (three dimensional depths) of $L1$ is to be estimated. The Moho is replaced directly by the Moho discontinuity in CRUST1.0 (Laske et al., 2013), which is shown in Figure \ref{fig4}b. 

3. $z_0$ and $k_0$

$z_0$ is a key parameter for the inversion geometry of  $L1$. For the study area, $z_0=18.4\rm{km}$, which is the mean depth of the upper crust from CRUST1.0 (Laske et al., 2013).  

$k_0=3.0\rm{Wm}^{-1}\rm{K} ^{-1}$ (eg., Turcotte and Schubert, 2014).

\subsection{The geometry of $L1$ for the Ordos geological block and its adjacent areas}

We first take a look at the fitness between the surface HF observed (Figure \ref{fig3}a) and that calculated from the HP model obtained in this paper (Figure \ref{fig3}b).  It can be observed that they are almost the same. At most of the grids the relative error is less than 1\% (Figure \ref{fig3}c). The maximum relative error is 13\%. These above show that the HP model obtained here satisfied the constraints of the surface HF. 

Figure \ref{fig4}a shows the depths or the topography of $L1$ ($1^\circ\times 1^\circ$) for the Ordos geological block and adjacent areas.  It can be found that this HP interface distributes unevenly. It is shallower in the south and deeper in the north part of the study areas. From Figure \ref{fig3} and Figure \ref{fig4}a it can be seen that this HP interface is shallower where surface HF is high, and deeper where surface HF is low. For the Ordos geological block, this HP interface is deeper in the interior than that in the boundary zones. On the other hand, this interface is shallower than the Moho discontinuity (Figure \ref{fig4}b), but there has no a clear relationship between them.

To show the effect of $z_0$ on the inversion results, Figure \ref{fig5}a and \ref{fig5}b shows the depth or tomography of $L1$ when $z_0$ has a change of $\pm 10.0$\% relative to that in Figure \ref{fig4}a, respectively. It can be observed that $L1$ shallow when $z_0$ decreases, otherwise $L1$ deepens, but the tendency of the tomography of $L1$ is similar to that in Figure \ref{fig4}a. From Figure \ref{fig5+1} we know that most of the relative error is about $\pm 10.0$\%. Therefore, how to determine a reasonable $z_0$, or how to present constraints on $z_0$ is important to the methodology here. 

At the end of this section, it should be noted that the geometry of $L1$ is probably one of the multi-solutions, as most of the inversion problems have. As pointed out in the previous, we do not know whether the variation of the surface HF is due to the distribution of the HP (sources), or geometry of the HP interface, or both. Here we tend to the geometry of the HP interface for simplicity, but we do not mean that there is no contribution from the other two cases. Therefore, our HP model for the Ordos geological block is only the one of the models that fit the observation of the surface HF.  So does the geotherm in the lithosphere based on this HP model for the Ordos geological block in the following. 

\begin{figure}[htb]
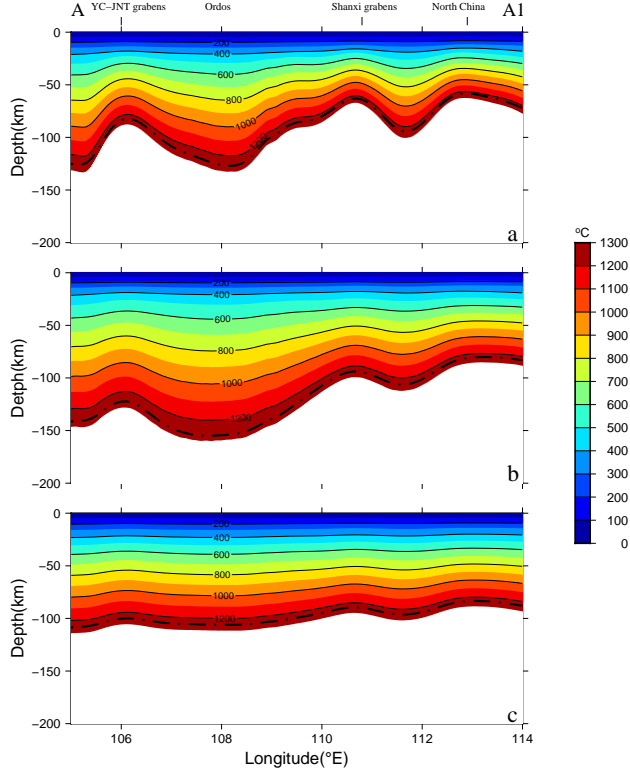

\setlength{\belowcaptionskip}{0pt}
\centering
\begin{overpic}[scale=0.4,bb=3 50 591 791]{Poumian_EW.eps}
\end{overpic}
\renewcommand{\figurename}{Fig.}
\caption{\footnotesize Geotherm for a cross-section AA1. Dash-dotted line is the bottom of the thermal lithosphere.}
\label{fig7}
\end{figure}

\section{Geotherms for the Ordos geological block and its adjacent areas}\label{sec4}

With the HP model obtained here (hereafter we call it Model (a)), we can study the geotherms for the Ordos geological block and its adjacent areas. For comparison, a layered HP model similar to Hasterok and Chapman (2012) (Model (b)) in which the bottom of the HP upper crust is the same to that from the CRUST1.0 (Laske et al., 2013), and a modified exponential decreasing HP model (Zang et al., 2002, 2005)(Model (c)) will be used. The procedure for calculating the gerotherm is similar to Zang et al. (2002, 2005) and Hasterok and Chapman (2011).

\begin{figure}[htb]
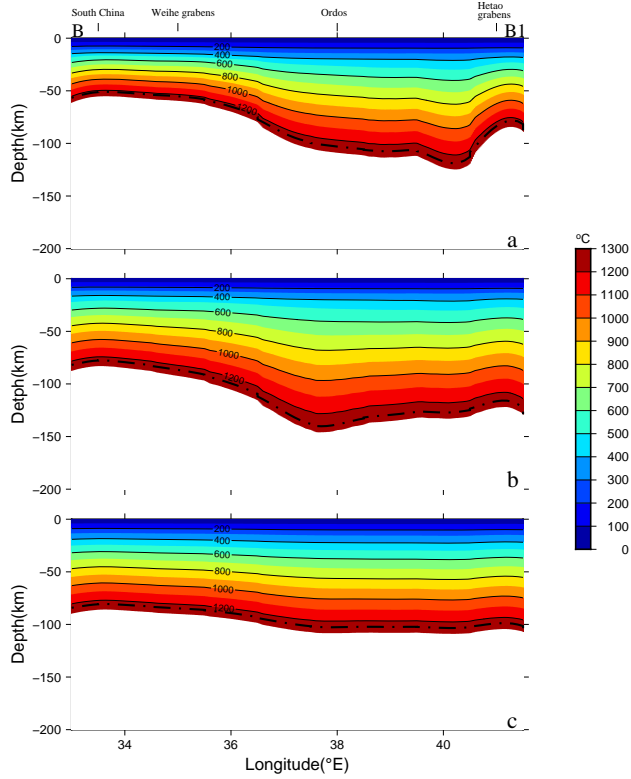

\setlength{\belowcaptionskip}{0pt}
\centering
\begin{overpic}[scale=0.4,bb=3 50 591 791]{Poumian_SN.eps}
\end{overpic}
\renewcommand{\figurename}{Fig.}
\caption{\footnotesize Geotherm for a cross-section BB1. Dash-dotted line is the bottom of the thermal lithosphere.}
\label{fig8}
\end{figure}

\subsection{1-D geotherm for two sites}

Figure \ref{fig6}a and b show the one-dimensional geotherm for S1 ($112.5^\circ\rm{E}, 40.5^\circ\rm{N}$) and S2 ($105.5^\circ\rm{E}, 34.5^\circ\rm{N}$), respectively.  It can be observed that Model (a) has somewhat higher temperature than Model (c),  while the latter has higher temperature than Model (b).  In Figure \ref{fig6}b and at $z=100 \rm{km}$, the temperature is about $1200^\circ\rm{C}$ in Model (a) but about $1000^\circ\rm{C}$ in Model (b).  Also plotted on this figure are the cenozoic xenolith P-T points near these two sites.  All geotherms of three models come close to matching the xenolith P-T points in Figure \ref{fig6}a, but Model (b) produces lower geotherm than that defined by the xenolith P-T points in Figure \ref{fig6}b. If the xenolith P-T points are assumed to represent present-day thermal conditions, the geotherm from Model (a) is reasonable.

For model (a), its geotherm is also reasonable even when $z_0$ has a change of $\pm 10$\% relative to that in Figure \ref{fig4}. When $z_0$ increases, $L1$ deepens, and temperature decreases; Otherwise, $L1$ shallows, and temperature increase. The detail change of the temperature at some different depths can be found in Table \ref{tb1}.

\begin{table}[htbp] 
\centering\footnotesize
 \begin{threeparttable}
  \caption{\label{tb1}{Temperatures at some depths for different HP Model (a) at S1 and S2.}}
 \begin{tabular}{cllllllllll}
 \toprule 
     & \multicolumn{5}{c}{S1} & \multicolumn{5}{c}{S2}\\
     \cline{2-11}
     Depth(km)& $T_0$($^\circ$C)&$T_0^-$($^\circ$C)&error(\%)& $T_0^+$($^\circ$C)&error(\%)& $T_0$($^\circ$C)&$T_0^-$($^\circ$C)&error(\%)& $T_0^+$($^\circ$C)&error(\%)\\
 \midrule
   10.0 &  212.0 &  212.0  &   0.0 &  212.0 &    0.0 &  218.6 &  219.0  &   0.2 &  218.6 &    0.0 \\
  20.0 &  387.8 &  391.7  &   1.0 &  385.1 &   -0.7 &  438.1 &  449.3  &   2.6 &  429.0 &   -2.1 \\
  40.0 &  624.6 &  641.2  &   2.7 &  608.5 &   -2.6 &  732.3 &  759.6  &   3.7 &  707.3 &   -3.4 \\
  60.0 &  836.0 &  865.6  &   3.5 &  806.0 &   -3.6 &  966.0 & 1008.0  &   4.3 &  926.1 &   -4.1 \\
  80.0 & 1042.2 & 1082.5  &   3.9 & 1000.1 &   -4.0 & 1176.3 & 1228.3  &   4.4 & 1125.2 &   -4.3 \\
 100.0 & 1232.4 & 1280.2  &   3.9 & 1181.5 &   -4.1 & 1364.3 & 1422.5  &   4.3 & 1305.9 &   -4.3 \\

 \bottomrule 
\end{tabular} 
\footnotesize Note: $T_0$: Temperature when $z_0$ is that in Figure \ref{fig4}a;$T_0^+$, $T_0^-$: Temperature when $z_0$ has a change of $\pm10$\% relative to that in Figure \ref{fig4}a, respectively. S1 ($112.5^\circ\rm{E}, 40.5^\circ\rm{N}$); S2 ($105.5^\circ\rm{E}, 34.5^\circ\rm{N}$.)
 \end{threeparttable} 
\end{table}

\subsection{Geotherm for two cross-sections}

Figure \ref{fig7}  and \ref{fig8} show the two-dimensional geotherm for two cross-sections of AA1 and BB1 in Figure \ref{fig2}, respectively.  It can be seen from Figure \ref{fig7} that Model (c) has higher temperatures than Model (a) and Model (b),  while Model (a) has somewhat higher temperatures than Model (b).  For example, in Ordos geological block, at about depth of 70 km, the temperature is about $1000^\circ\rm{C}$ in Model (c), but this temperature is at about depth of 80 km in Model (a) and 100 km in Model (b). The maximum depth for the bottom of the thermal lithosphere is at about 100 km in Model (c), but at about depth of 130 km in Model (a) and 150 km in Model (b).

It can also be seen from Figure \ref{fig7} that geotherms from Model (a) define a more reasonable trendency in tectonics than Model (b) and Model (c). The temperatures are high and have a dramatic variation in the grabens (YC-JNT and Shanxi grabens) where the tectonics is active. However, the temperatures are relatively low and have a gentle variation in the Ordos geological block. Similar properties can be seen from the distribution of the bottom of the thermal lithosphere, also from Figure \ref{fig8}. This tendency can also be clearly found from the geotherms of Model (b), but such a tendency is not obvious in geotherms of Model (c). 

\begin{figure}[htb]
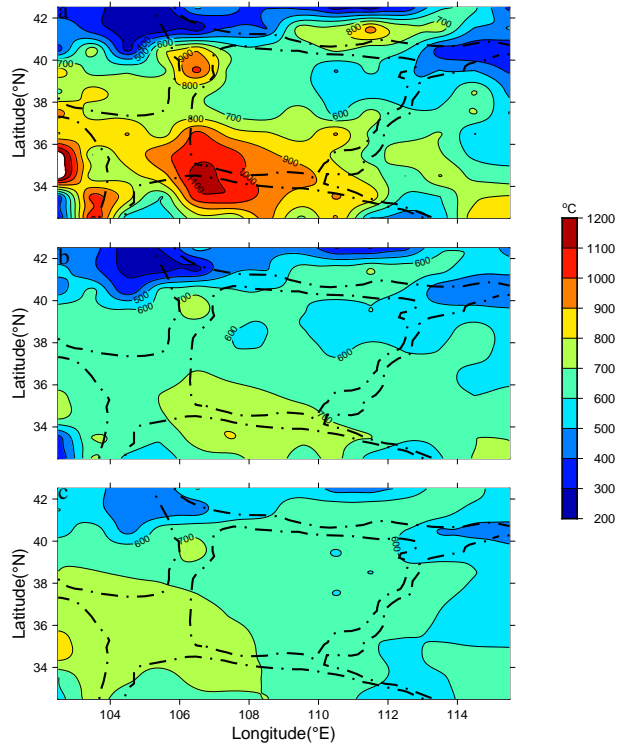

\setlength{\belowcaptionskip}{0pt}
\centering
\begin{overpic}[scale=0.4,bb=3 50 591 791]{T_moho.eps}
\end{overpic}
\renewcommand{\figurename}{Fig.}
\caption{\footnotesize Subfigure of a,b and c are temperatures at the Moho from Model (a), Model (b) and Model (c) , respectively.}
\label{fig9}
\end{figure}

\subsection{Geotherms at the Moho in Ordos block and its adjacent areas}

\begin{figure}[htb]
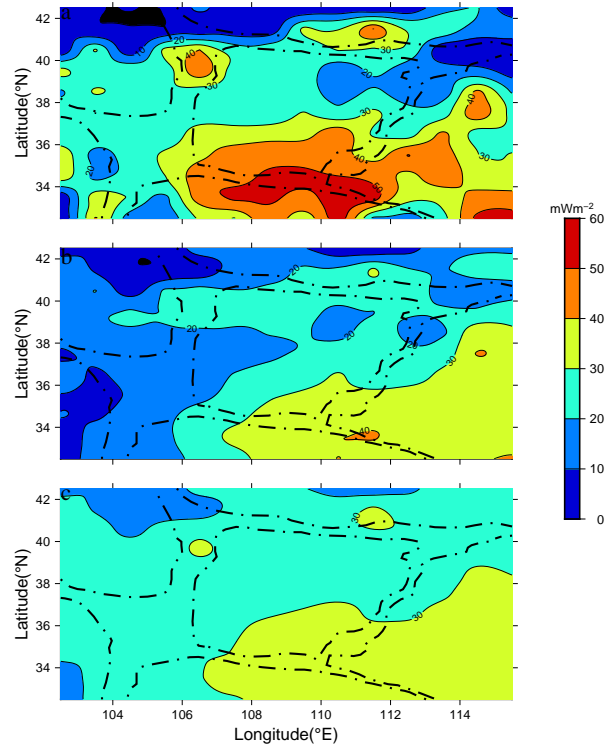

\setlength{\belowcaptionskip}{0pt}
\centering
\begin{overpic}[scale=0.4,bb=3 50 591 791]{Hf_moho.eps}
\end{overpic}
\renewcommand{\figurename}{Fig.}
\caption{\footnotesize Subfigure of a,b and c are heat flows at the Moho from Model (a), Model (b) and Model (c) , respectively.}
\label{fig10}
\end{figure}

Figure \ref{fig9}  shows the temperatures, and Figure \ref{fig10} heat flows, at the Moho from Model (a), Model (b) and Model (c) ($1^\circ\times 1^\circ$) , respectively.  Overall it can be seen that Model (a) has higher temperatures and HFs than Model (b),  and Model (c) has lower temperatures and HFs than Model (b); The temperatures and HFs at the Moho from Model (a) distribute unevenly: The southern and western part has relative higher temperatures and HFs than the northern and eastern part; Higher temperatures and HFs often distribute in the boundaries (grabens) and their adjacent areas, while the interior of the geological block has relative lower temperatures and HFs. These properties are clear in Model (a), then in Model (b).  

In Model (a), there are some areas where the temperatures at the Moho are less than $700^\circ\rm{C}$, and HFs are less than $20\mbox{\ mWm}^{-2}$. These areas may be the most stable. And there also are some active areas where the temperatures at the Moho are greater than $800^\circ\rm{C}$, and HFs are greater than $30\mbox{\ mWm}^{-2}$. The properties above are partly preserved in Model (b), but less in Model (c), which means that the temperatures in Model (b) and Model (c) are flattened, especially in the latter.  From the view of geotherm and this distribution trendency of temperatures and HFs at the Moho, Model (a) reflects most of the tectonics in the Ordos geological block and its adjacent areas.

In the end, it should be pointed out that the geotherm in Model (a) is almost dependent on the surface HF, because both the HP and HF boundary condition are from this observation. If the surface HF is reasonable, so does the geotherm.
This is different from the geotherm in Model (b) and (c), in which the geotherm seems to be smoothed and surface HF is not the dominant.

\section{Conclusions}\label{sec5}

We propose a method to infer the geometry of the HP layers in the stable continental lithosphere by inversion of the corresponding HF observation on the surface.  If the mean depth of the HP interface and the HP contrast between the two media are given, the geometry of the interface is iteratively calculated quickly. 

Based on Hasterok and Chapman's HP model for the continental lithosphere (Hasterok and Chapman, 2011), we infer the geometry of the bottom of the HP upper crust for Ordos geological block and its adjacent area in China. Most depths for this bottom are between $10\sim 20\rm{\ km}$. 

Base on the HP model we obtained, the geotherm for the Ordos geological block is calculated, and compared it with those from other two HP models. The geotherm based on our HP model fit better the constraints from the studies on the xenolith and tectonics. Most of the temperatures at the Moho are between $600\sim 700^\circ\rm{\ C}$, and the heat flows are between between $20\sim 40\rm{\ mWm}^{-2}$. 

Case study in Ordos geological block shows that our method has an important parameter, $z_0$, the initial mean value of the geometry for the inversion interface. Different $z_0$ will cause different HP model. Once $z_0$ is determined, only surface observation of HF required. One to three dimensional HP model can easily be obtained with this method. So this method is an alternative one for inferring the HP within the stable lithosphere.

\vspace{5em}

\def\thebibliography#1{
{\Large\bf  References}\list
 {}{\setlength\labelwidth{1.4em}\leftmargin\labelwidth
 \setlength\parsep{0pt}\setlength\itemsep{.3\baselineskip}
 \setlength{\itemindent}{-\leftmargin}
 \usecounter{enumi}}
 \def\newblock{\hskip .11em plus .33em minus -.07em}
 \sloppy
 \sfcode`\.=1000\relax}
\let\endthebibliography=\endlist

\end{document}